# Effective Hyper-clutter Artifacts Suppression for Ultrafast Ultrasound Doppler Imaging

Lijie Huang, Jingyi Yin, Jingke Zhang, U-Wai Lok, Ryan M. DeRuiter, Kaipeng Ji, Yanzhe Zhao, Tao Wu, James D. Krier, Xiang-yang Zhu, Andrew J. Bentall, Andrew D. Rule, Thomas D. Atwell, Lilach O. Lerman, Shigao Chen*, and Chengwu Huang*

*Abstract*—*Objective:* Hyper-clutter artifacts (HCA), arising from strong tissue reflections or physiological motion, present persistent challenges in ultrafast ultrasound Doppler imaging, often obscuring surrounding small vessel flow signals, especially in fascial regions such as the renal capsule. This study proposes *U-profile-based decluttering* (UPBD), a robust and computationally efficient method that exploits singular value decomposition (SVD)-derived spatial singular vectors to suppress HCA in ultrafast Doppler imaging. *Methods:* UPBD analyzes intensity profile of each pixel along the singular-order dimension of the SVD-derived left singular vectors *U*. A pixel-wise clutter-energy ratio is computed to derive a spatially adaptive declutter weighting map, which is applied to the SVD-filtered flow signals. *Results:* UPBD was evaluated on multiple *in vivo* datasets. Quantitative assessments based on contrast-to-noise ratio (CNR) and contrast-to-tissue ratio (CTR) demonstrated significant improvements over conventional SVD filtering. For example, UPBD enhanced CTR from 7.3 dB to 21.7 dB in contrast-free pig kidney, 17.8 dB to 42.1 dB in contrast-enhanced pig kidney, 8.2 dB to 32.8 dB in human kidney, and -4.9 dB to 3.7 dB in 3D human liver. *Conclusion:* The proposed UPBD method effectively suppresses HCA while preserving blood flow signals with minimal extra computational cost and no need for extensive parameter tuning. *Significance:* UPBD serves as a lightweight, easily integrated post-processing method that enhances HCA suppression, enabling broader application of SVD-based ultrafast Doppler imaging.

## I. Introduction

ULTRASOUND microvascular imaging (UMI), known as ultrafast Doppler, has recently emerged as an imaging modality that enables noninvasive visualization of small vessels and low-velocity blood flow without the need for ultrasound contrast agents [1-6]. It has been successfully applied to the assessment of microvascular structures in various organs, including the brain [7-9], kidney [10, 11] , and liver [12]. The development of UMI has been enabled by high-frame-rate imaging strategies, such as plane wave [13, 14] and diverging wave [15] transmissions with frame rate up to 10 kHz. Building on the principles of conventional Doppler ultrasound, ultrafast Doppler imaging leverages ultrafast acquisition and advanced clutter filtering methods to enhance the detection sensitivity of microvessels [16].

In conventional Doppler ultrasound imaging, a high-pass filter was typically used to suppress slowly varying tissue signals corresponding low frequency components in Doppler spectrum [17, 18]. However, the high-pass filter is ineffective in distinguishing tissue and low-velocity blood flow signals when their Doppler frequencies overlap, particularly in regions with complex tissue motion. Advancement in clutter filtering techniques, such as eigen based or singular value decomposition (SVD) based cluttering [19], has shown promise in improved suppression of clutters for blood flow detection. Demené et al. [16] applied SVD to ultrafast ultrasound imaging to achieve substantial sensitivity to small vessels. Unlike velocity-based filters, SVD exploits the spatiotemporal coherence of signals to separate tissue and blood flow. Nevertheless, cutoff rank selection remains largely heuristic or empirical, introducing subjectivity and potentially limiting the robustness and generalizability of SVD-based clutter filtering. To address this, widely adopted approaches in SVD-based UMI include curvature analysis of the singular value spectrum (elbow method) [20] and spatial similarity matrix (SSM)-based estimators [21, 22]. Beyond rank selection, several advances have been proposed to improve the practicality of SVD filtering. Randomized SVD reduces the cost of matrix decomposition [23], whereas block-wise or adaptive local SVD [20, 24] enables spatially adaptive filtering by adjusting thresholds to local motion and tissue properties. High-order SVD (HOSVD) further extends decomposition into a tensor domain [25], integrating higher-order information such as channel dimension. However, both local SVD and HOSVD substantially increase computational burden, which limits their real-time implementation.

Although clutter filtering for blood flow imaging has attracted considerable attention, residual hyper-clutter artifacts (HCA) remain a challenge [26], particularly in regions such as

Research reported in this publication was partially supported by the National Institute of Diabetes and Digestive and Kidney Diseases and the National Institute of Arthritis and Musculoskeletal and Skin Diseases under Award Numbers of R01DK129205, R01DK138998, and R21AR076028. The content is solely the responsibility of the authors and does not necessarily represent the official views of the National Institutes of Health. The Mayo Clinic and some of the authors (L. H., C. H., and S.C.) have pending patent applications related to the technologies referenced in this publication.

(Corresponding authors: Shigao Chen and Chengwu Huang).
L. Huang, J. Yin, J. Zhang, U.-W. Lok, R.M. DeRuiter, K. Ji, Y. Zhao, T. Wu, T.D. Atwell, *S. Chen and *C. Huang are with the Department of Radiology, Mayo Clinic College of Medicine and Science, Rochester, MN 55905 USA (e-mail: Chen.Shigao@mayo.edu; Huang.Chengwu@mayo.edu).
J. D. Krier, X.Y. Zhu, A.J. Bentall, A.D. Rule, L. O. Lerman are with the Division of Nephrology and Hypertension, Mayo Clinic, Rochester, MN 55905 USA.

capsules, fascia, and vessel walls. In SVD clutter filtering, these components can leak into blood-dominated ranks, especially when there is considerable tissue motion, making it difficult to be suppressed completely. To address this issue, Wang et al. [26] introduced the concept of hyper-motion scattering (HMS) and proposed an adaptive SVD filtering method. Specifically, HMS and non-HMS regions in pre-beamformed in-phase quadrature (IQ) data were first identified using k-means clustering, after which region-specific SVD filtering was applied to selectively suppress HMS artifacts. This approach demonstrated significantly improved suppression of HMS in human carotid artery imaging across different cardiac phases, although it requires access to pre-beamformed data.

Despite these advances, most SVD-based clutter suppression strategies remain centered on singular value spectra or the SSM derived from spatial singular vectors ($U$), whereas other informative characteristics of $U$ have not been systematically explored. The $U$ matrix represents the spatial distribution of each singular component and pixels corresponding to tissue clutter, blood flow, or noise often exhibit distinct behaviors along the singular-order dimension in the $U$ matrix. In particular, the intensity profile of $U$ for each pixel along singular order (referred to as $U$-profile), providing complementary features for signal separation beyond singular values. Building on this idea, Huang et al. introduced an adaptive multilevel thresholding method that exploits spatial singular vector curve to determine cutoff ranks locally for SVD-based clutter filtering [27]. Developed for transthoracic coronary flow imaging, the method focuses on local adaptive thresholding rather than HCA suppression.

In this work, we propose $U$-profile-based decluttering (UPBD), an HCA suppression strategy that leverages the $U$ profiles (the intensity distribution per pixel of the left singular vectors $U$ across singular orders) to suppress HCA and enhance the visibility of blood flow. This enables adaptive suppression of strong reflectors and residual tissue clutter while preserving blood flow visibility. To demonstrate the generalizability of this strategy, we validate its performance across multiple *in vivo* datasets, including contrast-free and contrast-enhanced pig kidney, contrast-free human kidney, and 3D human liver. The proposed approach has the potential to be seamlessly integrated into ultrafast Doppler imaging, providing a practical and effective solution for improving HCA suppression in clinical applications.

## II. METHODS

The proposed UPBD method is built upon SVD. The overall workflow consists of several main steps. First, SVD is applied to the beamformed ultrafast ultrasound data to obtain the left singular vector $U$ matrix, and the cutoff rank is automatically determined. Next, $U$-profiles are constructed from $U$ matrix, representing the intensity profile of each spatial pixel across the singular-order dimension. These $U$-profiles are subsequently used to generate a pixel-wise declutter weighting map that adaptively suppresses HCA. Finally, the blood flow data are reconstructed from the SVD components and multiplied by the expanded weighting map to produce the final decluttered power Doppler image.

### A. Singular Value Decomposition

**Decomposition**: SVD-based clutter filtering is widely used for clutter filtering in ultrasound Doppler imaging due to its strong ability to separate tissue and blood flow components according to their distinct spatiotemporal coherence. The principle of the decomposition step is illustrated in Fig. 1a. Let the beamformed and compounded ultrafast ultrasound data be denoted as a 3D matrix $S \in \mathbb{C}^{M \times N \times T}$, where $M$, $N$, and $T$ represent the number of axial, lateral, and temporal samples, respectively. The data are first reshaped into a 2D matrix $\tilde{S} \in \mathbb{C}^{MN \times T}$, where each column corresponds to a spatially vectorized frame. The singular value decomposition of $\tilde{S}$ is given by:

$$\tilde{S} = U\Lambda V^* = \sum_{i=1}^{T} \lambda_i u_i v_i^* \quad (1)$$

Here, $U \in \mathbb{C}^{MN \times T}$ contains the spatial singular vectors, where each column $u_i$ represents a spatial basis associated with the $i$-th singular order. $V \in \mathbb{C}^{T \times T}$ contains the temporal singular vectors, with each column $v_i$ capturing the corresponding temporal variation. $\Lambda \in \mathbb{R}^{T \times T}$ is a diagonal matrix whose entries $\lambda_i$ are the singular values arranged in descending order. Tissue signals are primarily captured by the first few singular vectors associated with the largest singular values, whereas blood flow signals are represented in the higher-indexed singular vectors corresponding to smaller singular values.

**Cutoff Determination:** We determine the cutoff from the SSM computed on the spatial singular vectors. Specifically, let $U = [u_1, u_2, ..., u_T] \in \mathbb{C}^{MN \times T}$ denote the matrix of spatial singular vectors. The SSM is defined as,

$$SSM(i,j) = corr(|u_i|, |u_j|) \quad (2)$$

where $|u_i|$ represents the magnitude distribution of the $i$-th singular vector. In the SSM, tissue and blood subspaces typically appear as two highly correlated, square-like blocks aligned along the diagonal, as illustrated by the red boxes in Fig. 1a. Two adjacent diagonal blocks delineating the coherent regions are identified [22] and the boundary between them is defined as the SVD cutoff rank $L_1$ separating tissue from blood flow.

**Reconstruction**: The filtered data $\tilde{S}_{blood}$ is reconstructed as:

$$\tilde{S}_{blood} = \sum_{i=L_1}^{T} \lambda_i u_i v_i^* \quad (3)$$

where $L_1$ is the determined SVD cutoff rank. The matrix $\tilde{S}_{blood}$ is then reshaped into a 3D format $S_{blood} \in \mathbb{C}^{M \times N \times T}$ to generate PD image (Fig. 1c).

### B. U-Profile-Based Decluttering

In the $U$-profile-based analysis, each singular vector $u_i$ was reshaped into a 2D spatial map of size $M \times N$ for visualization (Fig. 1b). For each spatial pixel $j \in [1, MN]$, we define its $U$-profile as the intensity vector formed by the $j$-th row of the left singular vector matrix $U$:

$$\mu^j = \left[ |u_1^j|^2, |u_2^j|^2, ..., |u_T^j|^2 \right] \in \mathbb{C}^T \quad (4)$$

Here, $u_i^j$ denotes the $j$-th row element of the $i$-th left singular vector $u_i$, and $T$ is the ensemble length. This profile characterizes the distribution of signal intensity at pixel *j* across singular orders, thereby capturing its relative contribution to tissue and blood subspace.

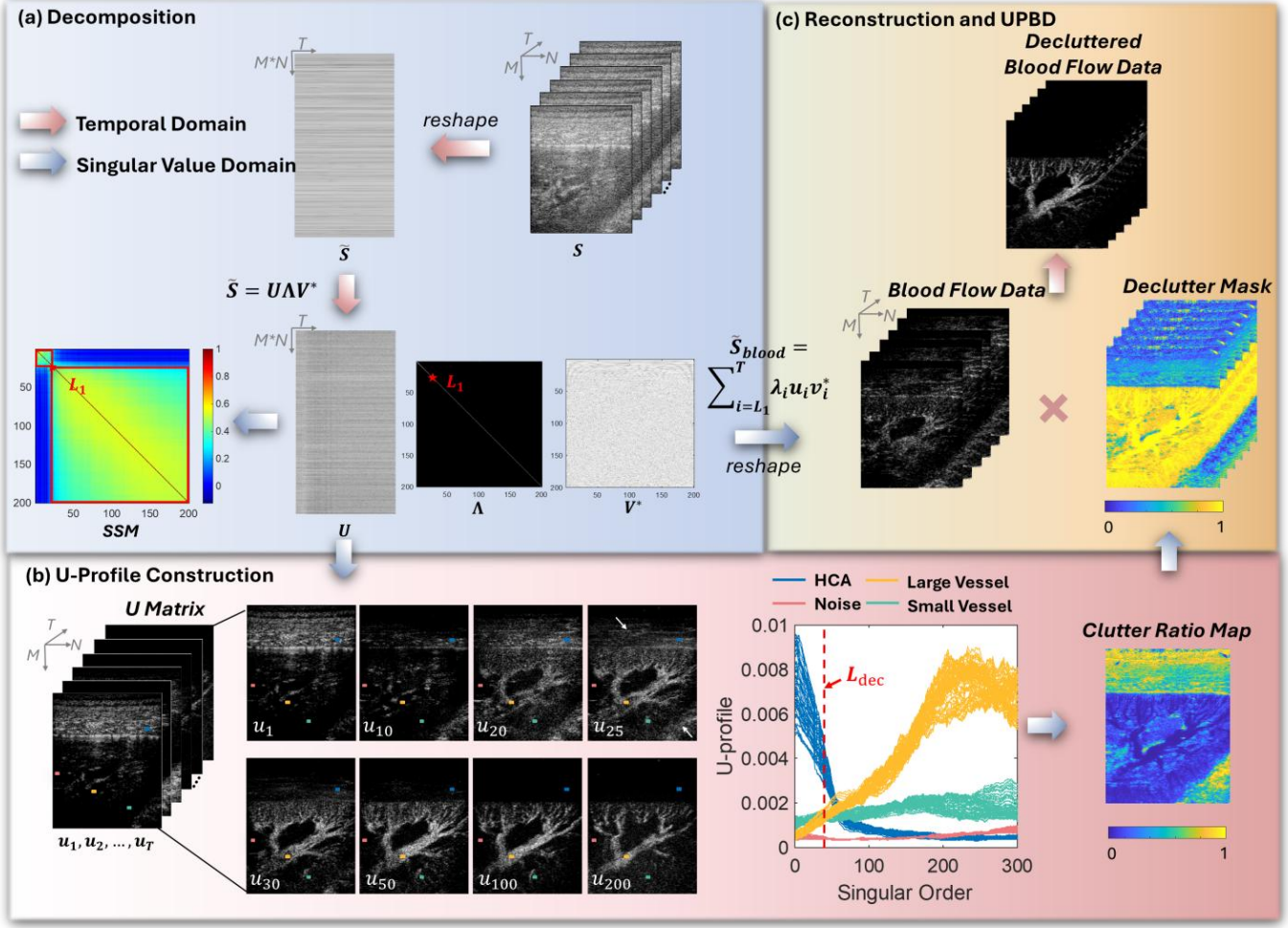

Fig. 1 Overview of the proposed *U*-Profile-based Power Doppler (UPBD) method. (a) Decomposition of the reshaped Casorati matrix with cut-off rank selection using the SSM method. (b) *U*-profile analysis of left singular vectors, which has different pattern between HCA, blood flow, and noise, and generates a clutter ratio map. (c) Reconstruction of blood flow data with application of the declutter mask, yielding the final UPBD output with reduced hyper-clutter and preserved microvascular signals.

In detail, the middle panel of Fig. 1b illustrates the *U*-profiles extracted from representative regions of HCA (blue), large vessels (yellow), small vessels (green), and noise (pink), corresponding to the colored boxes marked in the *U* matrix in the left panel. These *U*-profiles show clearly differentiated patterns across the singular-order dimension: HCA exhibits high intensity in the lowest orders followed by a rapid decay; large vessels display stronger contributions in the mid-to-high orders; small vessels present weaker and smoother profiles with modest mid-order elevations; and noise shows a relatively uniform distribution with a gradual rise toward higher orders. These observed differences motivate the construction of a pixel-wise weighting map for HCA suppression.

To quantify this distinction, we compute a clutter ratio map $R_{\text{clutter}}$, which calculated clutter–energy ratio for each pixel *j* as:

$$R_{\text{clutter}}(j) = \frac{\sum_{k=1}^{L_{dec}} |u_i^j|^2}{\sum_{k=1}^{T} |u_i^j|^2} \quad (5)$$

where $L_{dec}$ denotes the declutter cutoff, which could be defined as the optimal SVD cutoff rank obtained from the initial filtering using the SSM method. Pixels dominated by HCA typically yield larger $R_{\text{clutter}}$, whereas vessel-related and noise components exhibit smaller values because their *U*-profiles are less concentrated in the low-order components, as shown in the right panel of Fig. 1b.

Based on this ratio, we construct a pixel-wise declutter weighting map:

$$Dec(j) = 1 - R_{\text{clutter}}(j) \quad (6)$$

The 2D clutter ratio map is then expanded over time to match the size of the reconstructed blood-flow data $S_{blood} \in \mathbb{C}^{M \times N \times T}$, and then applied pixel-wise to $S_{blood}$ as shown in Fig. 1c.

C. Data Acquisition and Preprocessing

All *in vivo* ultrasound data were acquired using a Verasonics Vantage 256 system (Verasonics Inc., Kirkland, WA, USA) equipped with a GE9L-D linear array probe with a pitch of 0.23 mm (GE Healthcare, Wauwatosa, WI, USA) for 2D acquisitions and a 2D matrix array transducer (Vermon S.A., Tours, France; nominal 3 MHz) for 3D imaging. The datasets encompass a range of anatomical structures and experimental settings, including contrast-free and contrast-enhanced pig kidney, contrast-free human kidney and contrast-free human liver. All datasets were acquired using ultrafast plane wave imaging.

The pig study was approved by the Institutional Animal Care and Use Committee (IACUC) of Mayo Clinic. Pig kidney data were acquired from two domestic pigs, including one contrast-free and one contrast-enhanced acquisition. The domestic pigs were anesthetized with intramuscular Telazol (5 mg/kg) and Xylazine (2 mg/kg) [28]. For the contrast-free acquisition, the transmit center frequency was 5.208 MHz with a sampling frequency of 20.832 MHz. Ten plane waves were transmitted at steering angles from −9° to 9° with 2° increments. The pulse repetition frequency (PRF) was 5 kHz, and the frame rate was 500 Hz after compounding, and 200 frames of IQ data were collected over 0.4 s. The one-sided transmitted voltage was 50 V. For contrast-enhanced acquisitions, Definity microbubble (MB) suspension (Lantheus Inc., MA, USA) was administered as a 0.75 ml bolus via the external jugular vein, followed by a saline flush. This acquisition used the same center frequency, PRF, frame rate, and acquisition time as the contrast-free acquisition, but employed eight plane waves from −7° to 7° (2° increments) and a one-sided transmit voltage of 10 V. The spatial resolution of the beamformed IQ data was 0.5 $\lambda \times \lambda$ (axial × lateral).

Human studies were conducted under the approval of the Institutional Review Board (IRB) of Mayo Clinic, with informed consent obtained from all participants. The human kidney and human liver were imaged from two healthy volunteers, respectively. For human kidney study, a 10-angle plane wave imaging scheme was used, with an angular increment of 1° between -4.5° to 4.5°, a PRF of 10 kHz, and an effective frame rate of 500 Hz. The transmit center frequency was 6.25 MHz and the sampling rate was 25 MHz. The one-sided transmit voltage was 50 V. The spatial resolution of the beamformed IQ data was 0.5λ × 0.5λ (axial × lateral). A packet of 200 compounded IQ data was acquired during 0.4s for each dataset.

For the 3D human liver study, volumetric ultrafast plane wave imaging was performed using a 2D matrix array (Vermon S.A., Tours, France; nominal 3 MHz). The probe consisted of 1024 elements arranged in a 32 × 32 grid with a 0.3 mm pitch, yielding an active aperture of 9.3 mm (lateral) × 10.2 mm (elevational). A UTA 1024-MUX adapter was used to multiplex the 1024 elements to the 256 available system channels. The transmit center frequency was 3.47 MHz and the sampling frequency was 13.89 MHz. Eight plane-wave steering angles were transmitted. Four receive sub-apertures were sequentially acquired for each angle, resulting in 32 firings per volume. The PRF was 8,333 kHz per firing, and coherent compounding across the 8 angles yielded a volumetric frame rate of approximately 260 Hz. The one-sided transmit voltage was set to 25 V. 300 volumes of ultrafast IQ data were acquired within 1.15 s. The pixel size of the reconstructed volumetric data was 0.8λ × 0.8λ × 0.8λ (axial × lateral × elevational), with λ = 0.44 mm.

For each acquired dataset, SVD clutter filtering was applied using an ensemble size corresponding to the total number of frames to extract the blood flow signals [16]. The SVD cutoff ranks were automatically determined using the SSM method [22]. For 2D acquisitions, clutter-filtered IQ data ($M \times N \times T$) were obtained after SVD-based clutter filtering. PD images were reconstructed by accumulating the power of the IQ signals along the temporal dimension:

$$PD(m,n) = \sum_{t=1}^{T}|IQ(m,n,t)|^2 \qquad (7)$$

For 3D acquisitions, clutter-filtered volumetric IQ data ($M \times N \times L \times T$) were processed in the same way, yielding a volumetric PD dataset:

$$PD(m,n,l) = \sum_{t=1}^{T}|IQ(m,n,l,t)|^2 \qquad (8)$$

For visualization, maximum-intensity projection (MIP) was performed along the elevational direction of the 3D PD volumes. Quantitative analysis was based on this MIP image.

### D. Evaluation Metrics

Quantitative evaluation of the proposed UPBD method and the compared conventional method was performed on the final PD images using two metrics: contrast-to-noise ratio (CNR), and contrast-to-tissue ratio (CTR). These metrics evaluate the ability of the proposed UPBD method to reduce HCA while maintaining vessel visibility.

The CNR quantifies the contrast between blood flow regions and HCA, and is defined as:

$$\text{CNR} = \text{sign}(\Delta S) \times 10 \times \log_{10}\left(\frac{|\Delta S|}{\sigma_{noise}}\right)[dB] \qquad (9)$$

where $\Delta S = \bar{S}_{blood} - \bar{S}_{HCA}$, with $\bar{S}_{blood}$ and $\bar{S}_{HCA}$ denote the mean intensities within the blood flow and HCA regions, respectively, and $\sigma_{noise}$ is the standard deviation of a selected noise region. The absolute value ensures that the numerator is non-negative. A negative CNR would indicate that the HCA intensity exceeds the blood flow signal. The CTR is defined as:

$$\text{CTR} = 10 \times \log_{10}\left(\frac{\bar{S}_{blood}}{\bar{S}_{HCA}}\right)[dB] \qquad (10)$$

and specifically measures the contrast between blood flow and HCA regions, without explicitly accounting for noise.

All regions of interest (ROIs) for blood flow, HCA, and noise were manually delineated with reference to anatomical structures on the PD images. ROI locations and examples are shown in Figs. 2 and 4.

### III. RESULTS

#### A. Contrast-free and Contrast-enhanced Pig Kidney

Fig. 2 presents PD images obtained with conventional method and the proposed UPBD method from contrast-free and contrast-enhanced pig kidneys. Compared with conventional method, UPBD could effectively suppress the HCA as indicated by the white dash circles. In addition, vessels adjacent to clutter regions that appear obscured in Fig. 2c become better visible in Fig. 2d, as highlighted by the white arrowheads. The enlarged

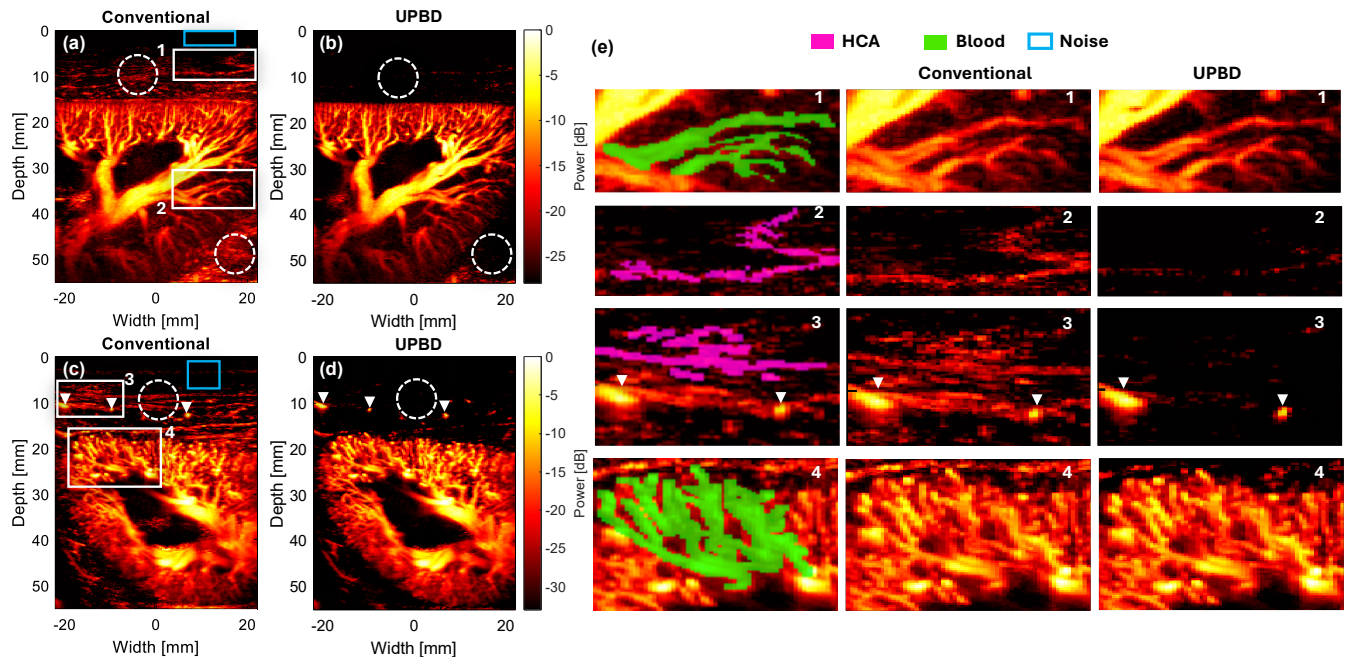

Fig. 2 Comparison of conventional power Doppler (PD) and the proposed UPBD method in *in vivo* pig kidney imaging under (a, b) contrast-free and (c, d) contrast-enhanced conditions. (a, c) Conventional PD images with regions of interest (ROIs) of HCA indicated by white boxes and noise regions marked by blue boxes. (b, d) Corresponding UPBD images. The UPBD results demonstrate effective suppression of residual HCA (white dashed circles) and improved separation of vessels from surrounding clutter compared with conventional PD (white arrowheads). (e) Enlarged ROIs from (a-d). The left panels show zoomed-in ROIs from (a) and (c), where HCA and blood flow are highlighted in magenta and green, respectively. The middle and right panels show the corresponding enlarged PD images obtained using the conventional method and the proposed UPBD method.

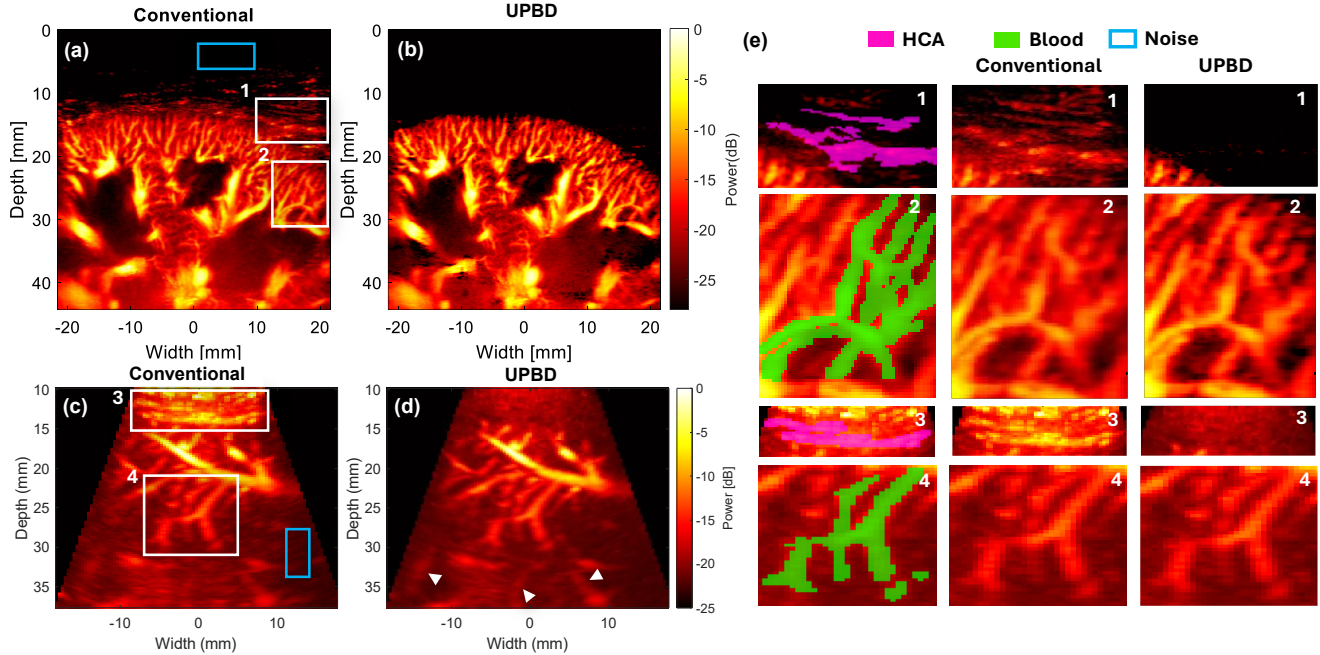

Fig. 3 Comparison of conventional PD and UPBD in *in vivo* human kidney and 3D human liver datasets. (a) Conventional and (b) UPBD power Doppler images for the human kidney. (c) Conventional and (d) UPBD MIP images for the 3D human liver. (e) Enlarged ROIs from (a-d). The left panels show zoomed-in ROIs from (a) and (c), where HCA and blood flow are highlighted in magenta and green, respectively. The middle and right panels show the corresponding enlarged PD/MIP images obtained using the conventional method and the proposed UPBD method.

ROIs (1–4) in Fig. 2e show representative regions from the conventional and the proposed methods. The left panels present the enlarged regions obtained with the conventional method, in which manually delineated ROIs for quantitative analysis are overlaid, highlighting HCA areas in magenta and blood flow regions in green. The middle and right panels present the corresponding enlarged ROIs obtained with the conventional and proposed methods, respectively.

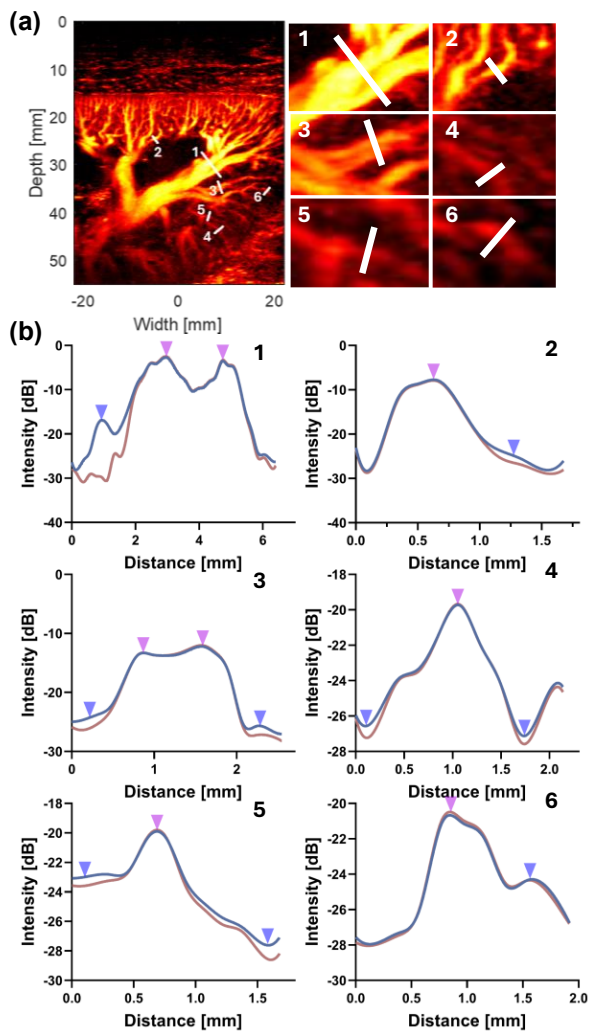

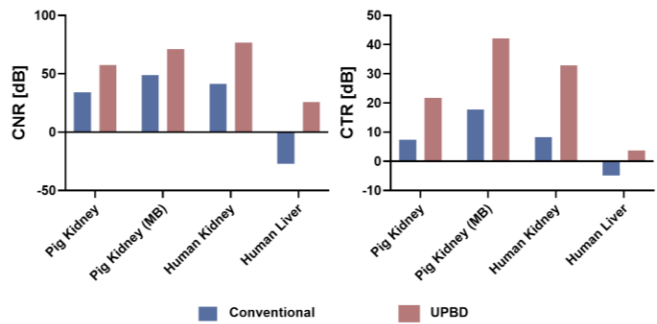

Fig. 4 Evaluation of blood flow and background intensities in contrast-free pig kidney PD images. (a) Six manually selected vessel cross-sections, including large vessels and small branches. (b) Intensity profiles of the selected vessels. Purple arrows indicate vessel peaks, where UPBD preserved flow signal amplitudes comparable to conventional PD. Blue arrows indicate background regions, where UPBD reduced noise.

To further evaluate whether UPBD preserves blood flow signals while suppressing HCA, we analyzed several vessel cross-sections in the contrast-free pig kidney dataset (Fig. 4). Six vessels of different sizes and intensities were manually selected (Fig. 4a), including both large vessels and small branches. The corresponding intensity profiles (Fig. 4b) show that the peak values at vessel locations (purple arrows) were comparable between UPBD and the conventional method, confirming preservation of blood flow signals. In the background noise regions around vessels (blue arrows), UPBD consistently reduced the intensity relative to the conventional method. These results demonstrate that UPBD not only suppresses HCA, as shown in Fig. 2, but also attenuates background noise around vessels, thereby improving vessel-to-background contrast.

**Fig. 5.** Quantitative evaluation of conventional PD and UPBD across pig kidney, pig kidney (MB), human kidney and 3D human liver data. Bar plots show (a) CNR and (b) CTR. Across both datasets, UPBD consistently outperforms the conventional method. Notably, in the human liver, the conventional method yields negative CNR and CTR values, indicating that HCA dominate over the blood flow signal, whereas UPBD achieves positive values, further demonstrating its effectiveness in suppressing HCA and enhancing vascular contrast. The ROIs used for quantitative analysis are indicated in Figs. 2 and 4.

Quantitative evaluation of the pig kidney datasets is summarized in Fig. 5. For both contrast-free and contrast-enhanced acquisitions, UPBD consistently achieved substantially higher CNR and CTR compared with conventional method, confirming its superior ability to enhance vessel visibility against HCA. This quantitative improvement validates the visual observations in Figs. 2 and 4, demonstrating that UPBD effectively suppresses HCA and improves vessel-to-clutter contrast.

*B. Contrast-Free Human Kidney and Liver*

Figs. 3 and 5 present the evaluation of UPBD on *in vivo* contrast-free human kidney and 3D human liver datasets. For the human kidney, conventional PD images were affected by HCA, as indicated by ROI 1 in Fig. 3a. By contrast, UPBD effectively suppressed HCA as shown in Fig. 3b. In the 3D human liver dataset acquired with the matrix probe, quantitative evaluation was performed on the MIP image. Compared with conventional method, UPBD markedly reduced HCA as shown in ROI 3 in Fig. 3d. UPBD improves the delineation of vascular structures even in the deep hepatic region where signal-to-noise ratio (SNR) is intrinsically low as indicated by the arrowheads in Fig. 3d. The enlarged ROIs in Figs. 3a and 3c are shown in Fig. 3e.

Quantitative results are summarized in Fig. 5. Across both kidney and liver datasets, UPBD consistently outperformed conventional method in terms of CNR and CTR. Notably, in the 3D human liver case, conventional method yielded negative CNR and CTR values, indicating that the HCA intensity exceeded that of the selected vascular regions. However, UPBD improved the CNR and CTR to positive values and substantially exceeded those obtained with the conventional method. This further highlights the advantage of UPBD in effectively suppressing HCA while preserving vascular signals, even in challenging 3D imaging scenarios.

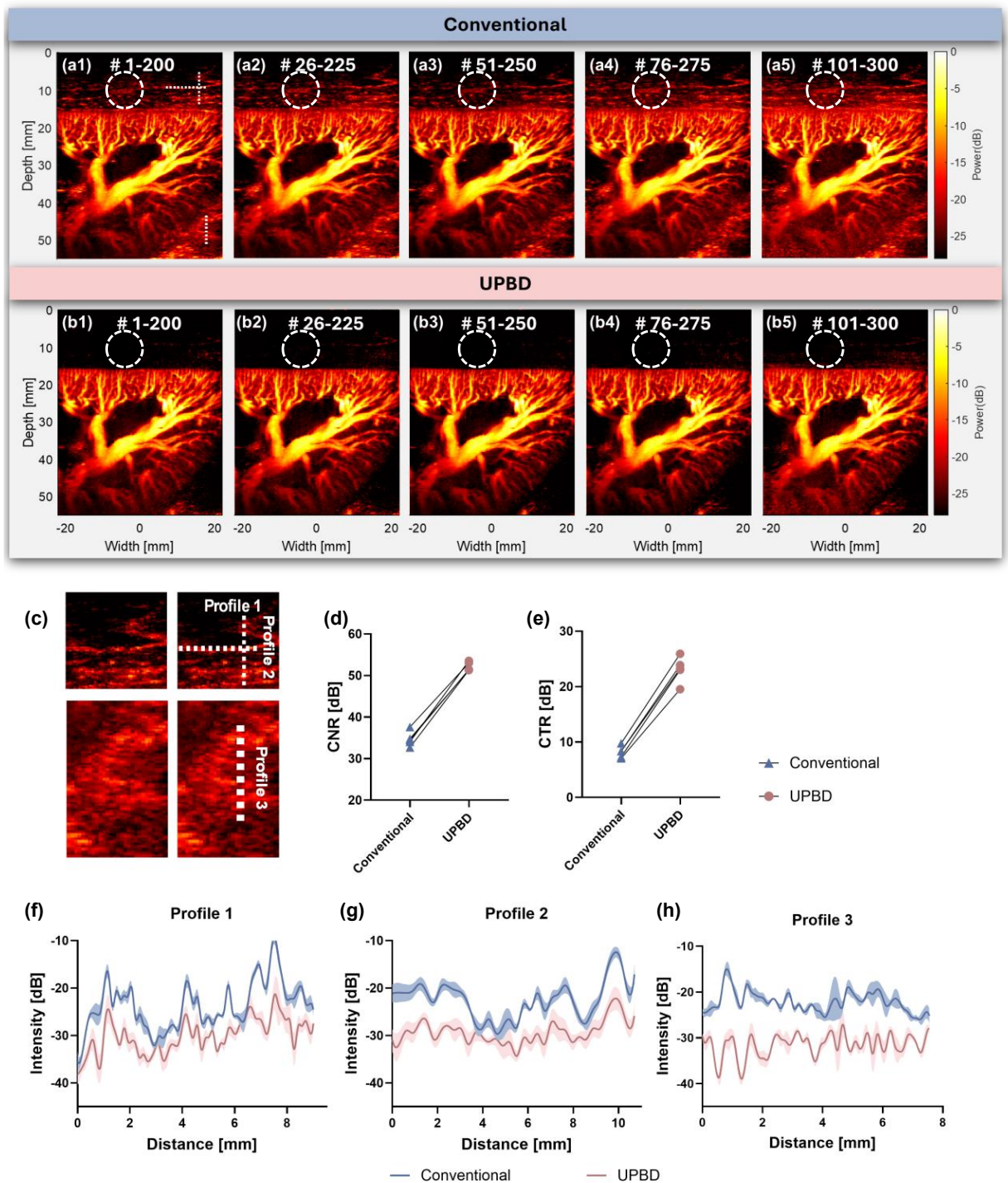

**Fig. 6.** Robustness of UPBD under sliding SVD processing in contrast-free pig kidney data. (a1–a5) Conventional PD images reconstructed with ensemble size = 200 and sliding step = 25 frames. (b1–b5) Corresponding UPBD results. (c) Zoomed-in view of three profiles extracted from HCA regions indicated in (c). (d–e) Quantitative evaluation of CNR and CTR across six sliding ensembles, computed on the same ROIs as in Fig. 2c. UPBD consistently outperforms the conventional method, with CTR showing the largest relative improvement (2–3 fold higher). (f–h) Ensemble-averaged HCA profiles with shaded regions representing standard deviation across ensembles. UPBD consistently yields lower HCA intensity and comparable variability with conventional PD, confirming its effectiveness and stability under different ensemble selections.

*C. Robustness on Sliding SVD*

The robustness of UPBD with respect to sliding SVD processing was evaluated using contrast-free pig kidney data. An ensemble size of 200 and a sliding step of 25 frames were used, producing five overlapping ensembles from the total 300-

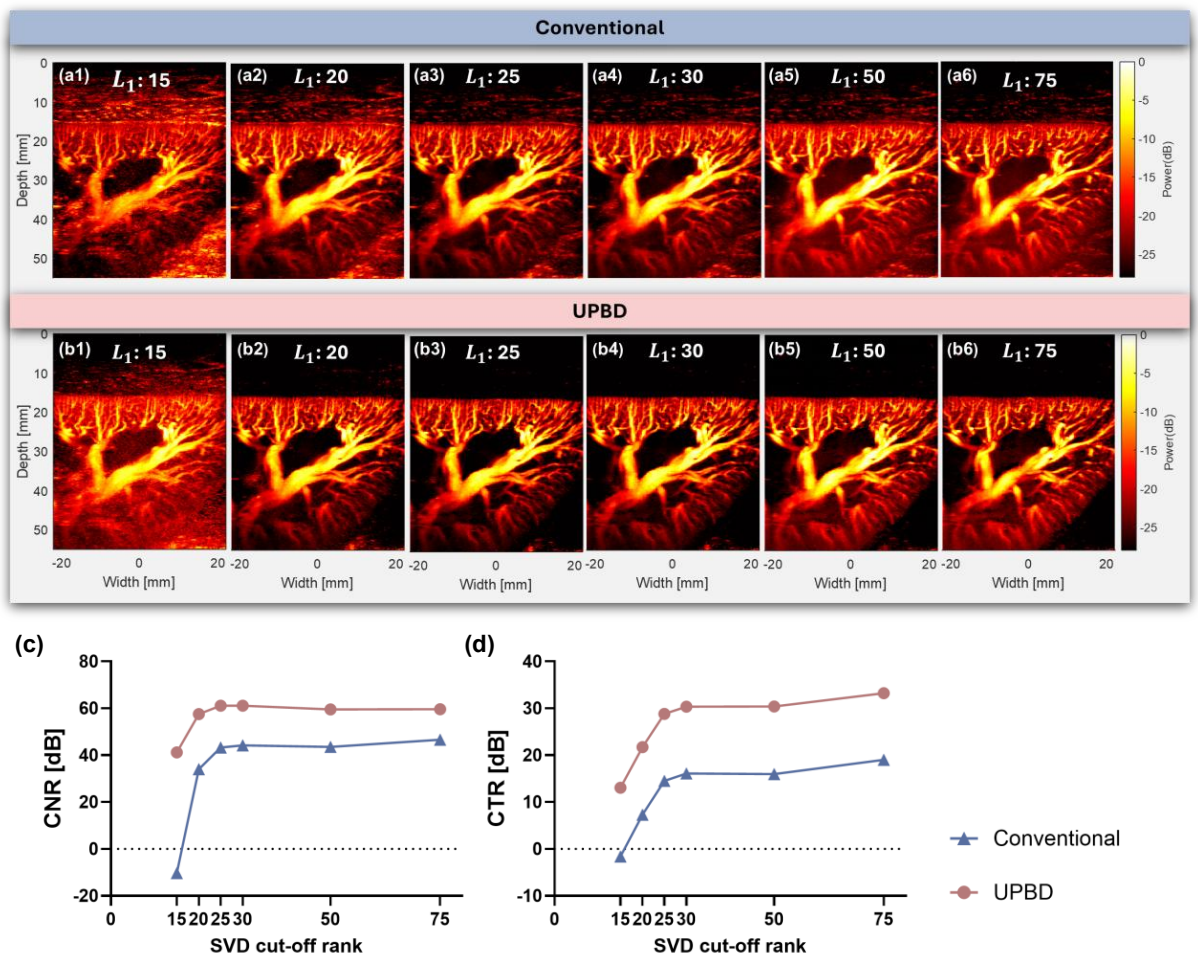

**Fig. 7.** Evaluation of UPBD performance across different SVD low cut-off ranks in contrast-free pig kidney data (ensemble size = 200). (a1–a6) Conventional PD results for ranks 15, 20, 25, 30, 50, and 75. (b1–b6) Corresponding UPBD results with the declutter mask applied at the optimal cut-off rank (20), applied to each SVD reconstruction. (c–e) Quantitative evaluation of CNR and CTR at increasing SVD low cutoff ranks. HCA decreases progressively with higher ranks, while UPBD consistently achieves additional suppression of HCA and background clutter. UPBD consistently outperforms the conventional method, demonstrating the robustness of UPBD to SVD rank selection.

frame dataset. For each ensemble, both the SVD cutoff rank $L_1$ and the declutter cutoff rank $L_{dec}$ were determined using the SSM method. Representative PD images reconstructed with the conventional method (Fig. 6a1–a5) and UPBD (Fig. 6b1–b5) demonstrate that UPBD consistently yields reduced HCA across all ensembles as indicated by the white dash circles. To further illustrate this effect, three-line profiles within HCA regions (indicated by dashed lines in Fig. 6a1) were analyzed. A zoomed-in view of the corresponding profiles is shown in Fig. 6c. Quantitative evaluation based on the ROIs defined in Fig. 2c is summarized in Fig. 6d–e. In all ensembles, UPBD outperformed the conventional method, with consistent gains in CNR and CTR. The improvement was most pronounced for CTR, which increased by a factor of 2–3 compared with conventional PD. The averaged line profiles across ensembles (Fig. 6f–h), with shaded regions denoting variability, further confirm that UPBD not only suppressed HCA more effectively but also yielded more stable results across sliding ensembles.

### D. Influence of Different SVD Cutoff Rank

To further investigate the effect of applying the declutter weighting map across different SVD cutoff ranks, we fixed the declutter cutoff $L_{dec}$ determined using SSM and varied the SVD low cutoff rank $L_1$. With an ensemble size of 200, $L_1$ of 15, 20, 25, 30, 50, and 75 were tested. The optimal low cutoff rank $L_1$ of SVD is calculated as 20 by the SSM method. The conventional PD results are shown in Figs. 7a1–a6, and the corresponding UPBD results in Figs. 7b1–b6. Qualitatively, HCA was progressively suppressed as SVD low cutoff rank $L_1$ increased, consistent with the expectation that higher-order singular vectors contain less clutter and more blood flow. At a rank below the optimal cutoff (e.g., 15), strong HCA and low-intensity background clutter remained visible, and applying the declutter weighting map was insufficient to fully suppress them. In contrast, for ranks above the optimal threshold (≥20), the UPBD mask effectively suppressed HCA indicated by the white dashed circle. The quantitative metrics in Fig. 7c–e further corroborate these findings: CNR and CTR all improved with increasing rank, with the steepest gains observed between ranks 15 and 20. These results confirm that UPBD maintains robust performance across different SVD reconstruction ranks and is particularly effective once the cutoff rank is above the optimal threshold. In contrast, conventional SVD filtering typically requires a much higher cutoff (around 75) to sufficiently remove HCA, which inevitably compromises the blood flow signal intensity, especially in the slow-flow regime. By comparison, UPBD achieved superior CNR and CTR at a cutoff of only 20, outperforming the conventional method at 75 according to the quantitative indexes. This indicates that using the optimal cutoff with UPBD not only suppresses clutter more efficiently but also preserves weak slow-flow signals, thereby enhancing microvascular sensitivity.

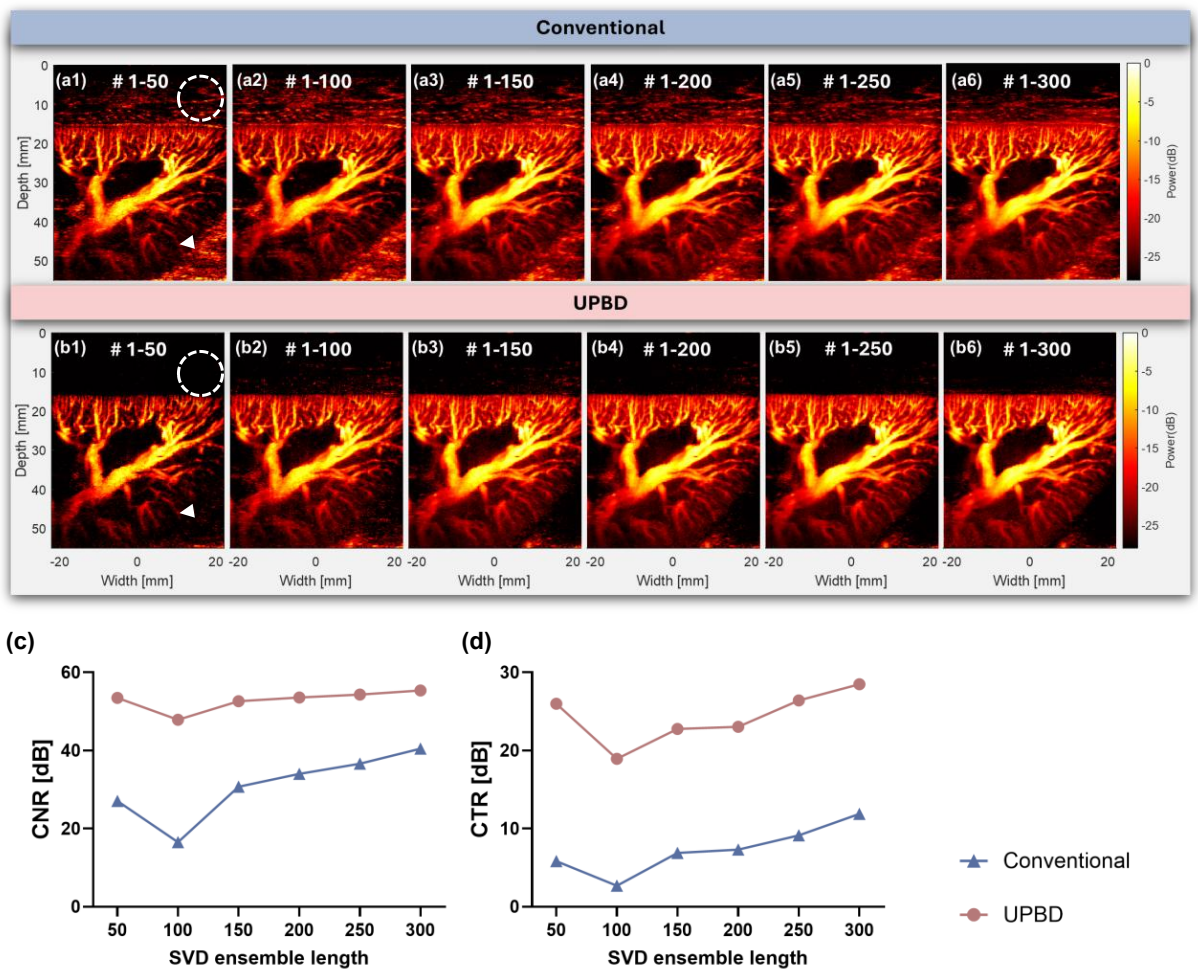

**Fig. 8.** Influence of SVD ensemble size on UPBD performance in contrast-free pig kidney data. (a1–a6) Conventional PD images with ensemble sizes of 50, 100, 150, 200, 250, and 300. (b1–b6) Corresponding UPBD results using the same ensemble-wise SVD cut-off ranks as in (a1-a6), *determined independently via SSM for each ensemble*. White dashed circles highlight the suppression of HCA by UPBD; however, white arrowheads reveal reduced visibility of deep small vessels when using a 50-frame ensemble. (c–e) Quantitative evaluation of CNR and CTR at increasing SVD ensemble size. Although the 50-frame ensemble yielded relatively high values, it was accompanied by suppression of small vessels. From 100 to 300 frames, all metrics improved steadily, demonstrating the benefit of larger ensembles for noise reduction, vascular preservation and contrast.

*E. Influence of Different SVD Ensemble Size*

The impact of ensemble size on UPBD performance was further investigated by varying the ensemble size from 50 to 300 frames. For each ensemble size, both the SVD cutoff rank $L_1$ and the declutter cutoff $L_{dec}$ were determined by the SSM method and set to the same value, ensuring consistent parameter selection across conditions. The conventional PD results are shown in Fig. 8a1–a6, and the UPBD results in Fig. 8b1–b6. Qualitatively, UPBD was able to suppress HCA even at the smallest ensemble size of 50, as indicated by the white dashed circle. However, as highlighted by the white arrowheads, suppression of deep microvessels was also observed at this short ensemble length. At an ensemble size of 100, vascular preservation improved, particularly for smaller vessels, while HCA suppression was maintained. Beyond 100, qualitative differences became less evident, although quantitative analysis (Fig. 8c–d) revealed important trends. At an ensemble size of 50, the quantitative metrics appeared relatively high. This counterintuitive result may result from insufficient separability between slowly moving tissue clutter and blood flow signals when the ensemble is too short. In this case, both clutter and part of the vascular signals were suppressed, leading to an artificial increase in contrast metrics despite the loss of small-vessel visibility. From 100 to 300 frames, CNR and CTR progressively increased, consistent with expectations that larger ensembles improve noise suppression and yield smoother vascular depiction.

## IV. Discussion

This study presents *U*-profile-based decluttering (UPBD), a simple and efficient strategy for suppressing hyper-clutter artifacts (HCA) in ultrafast blood flow imaging. By analyzing the rank-dependent intensity profile of spatial singular vectors, UPBD identifies HCA-dominated pixels and applies adaptive weighting directly to the clutter-filtered blood flow signals before PD reconstruction. In contrast-free pig kidney data, UPBD successfully suppressed HCA while preserving vascular signals (Fig. 4). Quantitative evaluations further demonstrated its superiority in contrast-enhanced pig kidney and human kidney datasets, where UPBD achieved substantial improvements in CNR and CTR compared with conventional PD imaging. Similar improvements were observed in the more challenging 3D liver dataset (Fig. 3). In conventional Doppler images, superficial regions exhibited strong HCA that exceeded the intensity of blood flow signals, while deeper vessels showed poor visibility due to attenuation. In contrast, UPBD effectively suppressed HCA and enhanced the visibility of deep vascular structures, resulting in a clearer and more balanced depiction of microvascular structures across depths. During robustness evaluations, UPBD maintained stable performance under sliding-SVD ensembles, varied SVD reconstruction ranks, and SVD ensemble sizes, consistently reducing HCA and improving vessel-to-background separation.

Unlike many advanced clutter filtering techniques for ultrasound microvessel imaging [20, 24-26], UPBD does not require local processing, pre-beamformed raw channel data, or high-dimensional information such as angular or channel dimensions. It also avoids iterative optimization and extensive parameter tuning. UPBD is computationally efficient, interpretable, and easy to implement. Rather than replacing the base pipeline, it enhances existing global SVD filtering workflows by applying a spatial weighting strategy based on the rank-dependent intensity profile of singular vectors. This design enables UPBD to generalize well across different datasets and acquisition conditions, while remaining fully compatible with conventional SVD-based pipelines for both 2D and 3D imaging.

Despite the effectiveness of UPBD in suppressing HCA, several limitations should be acknowledged.

First, the method relies on the *U*-profile of spatial singular vectors to compute adaptive and localized weights. In vessels with particularly slow flow, the energy may concentrate in low-rank components, potentially producing weights similar to those of HCA. This may lead to partial attenuation of slow-flow vessels during HCA suppression. Such ambiguity is more pronounced under low-frame-rate acquisition conditions, where the separability between slow flow and tissue motion is inherently reduced. Similarly, when the ensemble size used for SVD is too small, as shown in Fig. 8a1 and b1, the *U*-profile becomes less distinctive, which increases the risk of over-suppressing small vessels along with clutter. In this study, the weighting map was implemented in its simplest form as $Dec(j) = 1 - R_{\text{clutter}}(j)$. In principle, it can be generalized by introducing tunable parameters, such as $\alpha\epsilon[0,1]$ and $\beta > 0$, to regulate the suppression strength, e.g., $Dec(j) = (1 - \alpha \cdot R_{\text{clutter}}(j))^{\beta}$. Such an extension, or alternative nonlinear mappings of $R_{\text{clutter}}$ may provide additional flexibility in balancing HCA suppression and the preservation of slow vessels.

Second, the evaluation in this study was based on standard image-quality metrics including CNR and CTR, calculated from manually selected vascular and HCA regions. These ROIs were chosen based on anatomical landmarks and empirical contrast in the PD images, in the absence of ground truth for reference, and may lead to segmentation bias. Future work may focus on developing more robust and objective metrics to better assess performance across various conditions.

Third, in the sliding SVD experiments, the use of a small sliding step led to substantial frame overlap between adjacent ensembles. While this allowed smoother transitions, it may have underestimated variability across cardiac cycles. Longer acquisitions with increased or non-overlapping sliding intervals could provide a more rigorous evaluation of temporal consistency of the proposed method.

Fourth, this study focused on abdominal imaging (kidney and liver), where physiological motion due to respiration and cardiac pulsation can introduce significant artifacts. However, the use of ultrafast acquisition with short ensemble durations (approximately 0.5 s) helped mitigate motion-related effects in the present experiments. Future studies are needed to validate UPBD in other anatomical regions with more complex motion, such as the carotid artery [26, 27].

In summary, the proposed UPBD method provides a robust, interpretable, and computationally efficient solution for enhancing SVD-based ultrafast Doppler imaging. Its compatibility with existing pipelines, strong generalizability across datasets and imaging conditions, and minimal computational complexity make it a practical tool for improving UMI. Its simplicity and effectiveness pave the way for broader application in real-time and clinical ultrasound systems.

## V. Conclusion

This study presents a simple and clinically translatable UPBD method for HCA suppression in ultrafast Doppler imaging. The approach enhances Doppler image contrast by applying spatially adaptive weighting to the clutter-filtered blood flow data. Validated on contrast-free and contrast-enhanced datasets from both pig and human studies, including challenging 3D human liver imaging, the method consistently improves CNR and CTR. Its seamless integration into existing SVD pipelines and robust performance across imaging conditions highlight its potential as a scalable and clinically practical solution for enhanced microvascular visualization.